\documentclass[preprint,12pt]{elsarticle}

\usepackage{amssymb}

\usepackage{float}

\journal{arXiv}

\begin{document}

\begin{frontmatter}



\title{A SW Ka-Band Linearizer Structure with Minimum Surface Electric Field for the Compact Light XLS Project}


\author{M. Behtouei$^{1}$, L. Faillace$^{1}$, B. Spataro$^{1}$, A. Variola$^{1}$ and M. Migliorati$^{2, 3}$\\\vspace{6pt}}

\address{$^{1}${INFN, Laboratori Nazionali di Frascati, P.O. Box 13, I-00044 Frascati, Italy};\\
 $^{2}${Dipartimento di Scienze di Base e Applicate per l'Ingegneria (SBAI), Sapienza University of Rome, Rome, Italy};\\
  $^{3}${INFN/Roma1, Istituto Nazionale di Fisica Nucleare, Piazzale Aldo Moro, 2, 00185, Rome, Italy }}

\begin{abstract}
There is a strong demand for accelerating structures able to achieve higher gradients and more compact dimensions for the next generation of linear accelerators  for research, industrial and medical applications. 

In the framework of the Compact Light XLS project, an ultra high gradient higher harmonic RF accelerating structure is needed for the linearization of the longitudinal space phase. 
In order to determine the maximum sustainable gradients in normal conducting RF powered particle beam accelerators with extremely low probability of RF breakdown, investigations are in progress for using shorts  accelerating structures in the Ka-band regime. We here report an electromagnetic  design of a compact linearizer standing wave (SW) accelerating structure 8 cm long operating on $\pi$ mode, third harmonic with respect to the Linac frequency (11.994 GHz) with a 100 MV/m accelerating gradient and minimum surface electric field. Numerical electromagnetic studies have been performed by using the well known SuperFish, HFSS and CST computing software.
\end{abstract}



\begin{keyword}
Particle Acceleration, Linear Accelerators, Accelerator applications, Accelerator Subsystems and Technologies\end{keyword}

\end{frontmatter}


\section{Introduction}

The development of ever more advanced accelerating structures is one of the leading activity of the accelerator community. High gradients, efficiency and reliability of the accelerating structures have a strong and positive impact on linear accelerators projects. The next generation of linear accelerators requires unprecedented accelerating gradients for high energy physics and particle physics experiments. Technological advancements are strongly required to fulfill demands of compact linear accelerators for particle physics colliders, of new compact or portable devices for radiotherapy, mobile cargo inspections and security, biology, energy and environmental applications, and so on. The advantages of using high frequency accelerating structures are well known: smaller size, higher shunt impedance, higher breakdown threshold level and short filling time.

The room-temperature accelerating structures are being subject of intense research and development to reach higher gradient and reliability \cite{ref1,dolgashev2020materials}. Ultimately large electric gradients are also required for a variety of new applications, notably including the extreme high brightness electron sources for the FELs, RF photo-injector etc. Technological activities to design, manufacture and test new accelerating devices using different materials and methods are under way all over the world. To develop components with performances well beyond the existing  devices, it is fundamental to use significantly improved manufacture technologies and innovative designs. The Laboratori Nazionali di Frascati (INFN-LNF) is involved in the modeling, development and tests of RF structures devoted to particles acceleration with higher gradient electric field through metal device, minimizing the breakdown and the dark current. In particular, new manufacturing techniques for hard-copper structures are being investigated in order to determine the maximum sustainable gradients well above 100 MV/m and extremely low probability of RF breakdown. In the framework of the Compact light XLS project \cite{ref2}, funded  by the European Union's Horizon 2020 research and innovation programme under grant agreement No. 777431,  the main linac frequency is F = 11.994  GHz. In order to compensate the non-linearity distortions due to the RF curvature of the accelerating cavities, the use of a compact third harmonic accelerating structure working at F = 35.982 GHz is required \cite{ref3}.  
Since only the single bunch operation is foreseen, the beam dynamics is not affected by the long-range wake-fields and no dedicated dampers of the parasitic higher order modes are adopted for the linearizer structure. 
In refs. \cite{ref4,ref5,ref6,ref7,ref8,klystron1,klystron2}, it is shown that there are reasonable candidates for microwave tube sources which together, with RF pulse compressor, are capable of supplying the required RF power. The technologies in the Ka-Band accelerating structures, high power sources and modulators have also been developed by getting very promising results in order to reach a RF power output of (40-50) MW by using the SLED system.
RF stability during operation and tuning tolerances are important points for the RF structure design in the high frequency range. The complexity of machining, tight mechanical tolerances and alignments are therefore important aspects  which have to be taken into account in the design activity.
In the framework of the Compact Light XLS project, in order to obtain a longitudinal phase space linearization  we have designed  a possible compact third harmonic SW  accelerating structure operating at a frequency of  F= 35.982GHz  working on the $\pi$ mode at about 100 MV/m accelerating gradient. 
This report  discusses  the fundamental RF parameters dependence of a SW wave structure operating on $\pi$ mode as function of the iris's aperture and the cavity's geometry in order to get the optimized RF design. The figures of merit like the longitudinal shunt impedance, quality factor, coupling coefficient, transit time factor and so on, are calculated with the well known  Superfish, HFSS and CST numerical computing software.
Moreover, in order to estimate the breakdown thresholds, investigations on the modified Poynting vector and pulse heating are discussed. Finally operational limit due to RF breakdowns at higher power operation \cite{ref9}, \cite{ref10}, \cite{ref11} is discussed.

\section{Accelerating structure choice}

The design of compact particle accelerators for future generation is defined on the basis of a compromise among several factors: RF parameters, beam dynamics, narrow spread of electron energy, RF power sources, easy fabrication, small sensitivity to construction errors, economical reasons and so on. In order to minimize the input power requirements for a given accelerating gradient, the RF accelerating structures have to be designed with the aim of maximizing the shunt impedance. On the other hand, the accelerating section performances could be limited by effects such as the beam loading, instabilities, beam break-up etc., caused by the interaction between the beam and the surrondings.

As an example,  a figure of merit for the accelerating structure is the efficiency with which it converts average input electromagnetic energy per unit length, into average accelerating gradient. If  $P_b$ is the average beam power and $P_{rf}$ the average RF power fed into the structure, the small fraction of energy extracted by beam defined as  $\epsilon$ = $P_b / P_{rf}$  has to be kept well below to some per cent for getting a satisfactory energy spread.  
On the basis of these simple considerations, the global RF properties for designing the accelerating structure are therefore summarized and listed in the following:

1) High accelerating field gradient to reduce the accelerator length; 
2) High shunt impedance to reduce the requirement of RF power; 
3) Low ratios $E_p/E_a$ and $B_p/E_a$, where $E_p$ and $B_p$ are the peak surface electric and magnetic fields respectively, $E_a$ is the average accelerating electric field. The low ratios are required in order to achieve the highest possible field gradient before reaching the breakdowns condition estimated with the modified Poynting vector and the pulse heating effects and therefore to get a satisfactory thermal effects; 
 4) High ratio $E_a^2/W$ where $E_a$ is the average accelerating electric field, $W$ is the energy stored in the structure per unit length, that is a measure of the efficiency with which the available energy is used for the operating mode; 
5) Appropriate shape profile for avoiding the generation of  multipactoring phenomena which could limit the accelerating section performances \cite{ref12};
6) Low content of longitudinal and transverse wake-fields in order to reduce energy losses  and affect the beam dynamics.
The RF parameters of the accelerating structures scale with the frequency and this justifies the motivation of working in the Ka-band regime \cite{ref13}.  Moreover, in the design of high gradient accelerator compact structures well over 100 MV/m the voltage breakdown becomes a major constraint.
Our concern is to design an accelerating structure operating on the $\pi$ mode with the requirements referring to the Table 1.

No specific effect due to the beam loading and beam dynamics has to be expected since the operation with a small average current and single bunch is adopted. 

The third harmonic frequency of the main Linac implies small physical dimensions and thereby the dissipated power constitutes one of the main constraints.  A reasonable upper limit on the average power dissipation has been estimated to be around at 4 kW/m \cite{PV-B.Spataro}. To meet the full requirements by keeping a flexibility margin, a structure with simple geometry and of reliable construction with satisfactory mechanical tolerances has to be chosen. The detailed RF properties and the thermal behaviour of the working  $\pi$ mode are described in the following subsections.

\begin{table}[H]

\caption{Parameters list for the cavity design}

\begin{center}

\small{ \begin{tabular}{|| c| c||}
\hline
Design parameters&  \\ 
 \hline\hline
Frequency [GHz]&  35.982\\ 
\hline 
Average accelerating Electric Field [MV/m]  & 100\\ 
\hline
Axial length[cm]&8 \\ 
\hline
Iris Aperture diameter [mm]& 2 \\ 
\hline
 Ratio of phase to light velocity $(\nu_\phi / c)$ & 1 \\ 
 \hline
Pulse charge [pC]& 75\\ 
\hline
Rms bunch length ($\sigma_\tau (fs)$)& 350\\ 
\hline
Pulse repetition rate frequency [Hz]& 1-10\\
\hline
\end{tabular}}
\end{center}
\end{table}

\section{Accelerating structure design}
      
RF cavities are important components in any accelerator, and their most common use is to impart energy to the traversing particles. However, RF cavities have also been optimized for different purposes, not limited to accelerating particles, as it is requested for our application for the longitudinal phase space linearization. In addition, our aim is to have a short axial length of the RF structure for avoiding to affect the beam dynamics and for space reasons. Ultra-compact linear accelerators are known to require an operating frequency of 30-100 GHz. Higher operating frequency leads to higher accelerator gradients, with a corresponding smaller accelerator length. On the other hand, the knowledge of the electromagnetic field limit in RF structure is a major constraint. The cavity geometry has to be optimized and feasible in order to maximize the performance of the operation mode.
Preliminary estimations of the a  linearizer Ka-Band TW structure has already been discussed in other papers and presented at the First Annual Meeting of the Compact Light XLS project Workshop \cite{ref14,ref15,ref16}.
The SW Ka-Band linear structure is used in the Compact Light European project, for linearizing the longitudinal space phase in order to increase beam brightness. Extensive electromagnetic analysis of the structure has been performed for getting the optimized RF design. Much care was used in choosing the cavity shape in order to avoid breakdown effects at a high accelerating gradient. 

By assuming an iris thickness h = 0.667 mm \cite{ref14,ref15,ref16}, we here propose a possible short linearizer Ka-band SW structure working on $\pi$ mode and  quantify the influence of the iris radius and of its geometry  on the main RF parameters.  A similar procedure has been discussed in another paper for case of superconducting multicell accelerator \cite{ref17}. The RF structure design study has been performed by using SuperFish, HFSS and CST computing software \cite{ref18,ref19,ref20}. 

We decided to investigate a SW configuration operating on $\pi$ mode in order to get a satisfactory longitudinal shunt impedance and an acceptable iris aperture for practical beam dynamics considerations. The normal conducting structure works on the $TM_{010}$ mode where the longitudinal momentum is provided by on axis longitudinal electric field. The choice of the $\pi$ mode forces the cell lengh L to $\lambda$/2 at the work frequency.  A cell to cell coupling coefficient K of about (1-2) $\%$ is requested in order to achieve a field flatness and quick energy flow in the accelerating structure.
Figure 1 reports the investigated cavity with a circular iris shape, Figure 2 has an elliptic iris shape, and Figure 3 is the fillet iris shape.

\begin{figure}[H]
\begin{minipage}{11.5pc}
\includegraphics[width=10pc]{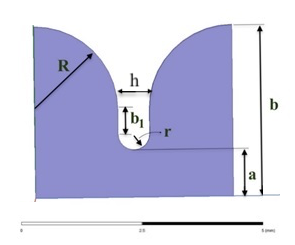}
\caption{\small{Cavity shape of the SW $\pi$ mode with circular iris shape. b, a, R, $b_1$ and r are cavity radius, iris aperture radius, cavity curvature radius, segment line and 1/2 of the iris thickness, respectively.}}
\end{minipage}\hspace{2pc}%
\begin{minipage}{8pc}
\includegraphics[width=10pc]{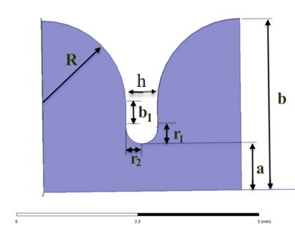}
\caption{\small{Cavity shape of the SW $\pi$ mode with a iris ellipse shape of $r_1$ and $r_2$ semi-axes shape. The other parameters are the same of Figure 1.}}
\end{minipage}\hspace{2pc}%
\begin{minipage}{9pc}
\includegraphics[width=10pc]{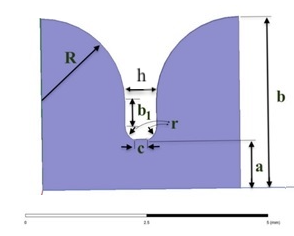}
\caption{\small{Cavity shape of the SW $\pi$ mode with a iris fillet shape. r and c are the iris fillet radius and segment between two fillets. The other parameters are the same of Figure 1.}}
\end{minipage} 
\end{figure}

In order to eliminate one point multipactoring \cite{ref12}, \cite{ref16} and to get a satisfactory low pulse heating effect, a cavity rounded shape profile is chosen. The cavity design has been performed by keeping a constant outer cavity radius R for different values of iris shapes and iris radius a. As a result, the maximum allowable iris radius is imposed by the outer radius R and iris shapes, in order  to avoid to cancel the segment line b$_1$ for keeping a proper cavity shape profile.
Electric and magnetic field distributions of the working $\pi$ mode are reported in Figs. $4a$ and Figure $4b$. The minimum value of the electric field and maximum value of the magnetic field are near the outer surface of the cavity as they were expected to be for the $TM_{010}$ working mode. 

\begin{figure}[H]
 \begin{center}
\includegraphics[width=0.4\linewidth]{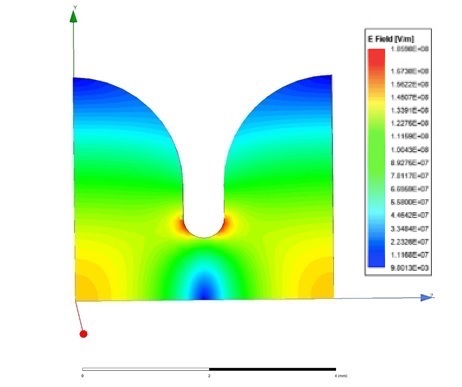}(a)
\includegraphics[width=0.4\linewidth]{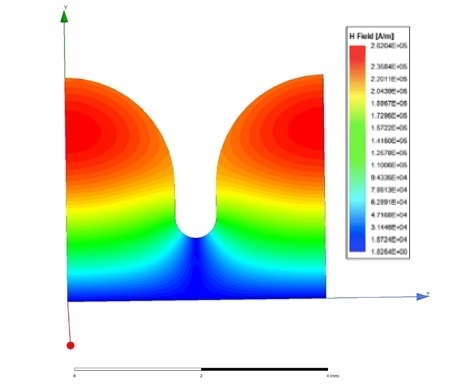}(b)

\caption { a) Electric field distribution of the $TM_{010}$, b) Magnetic  field distribution of the $TM_{010}$ mode }
\end{center}

      \end{figure}

Figures 5-9 illustrate the comparisons of the merit RF figures of the parameters shunt impedance $R_{sh}/m$, unloaded quality factor Q, coupling coefficient defined as K = ($F_\pi -F_0)/F_{\pi/2}$, where $F_\pi$, $F_0$ , $F_{\pi/2}$ denote the resonant frequencies of $\pi$, 0 and $\pi/2$ modes respectively, $E_p/E_a$ and $H_p/E_a$, as function of the $a/\lambda$ ratio with $\lambda$ = 8.331 mm for different iris shape by assuming a fixed iris thickness of 0.668 mm. The results are in agreement with the scaling laws of RF parameters with the frequency \cite{ref13}, \cite{ref21}.  In all cases, the fit of the  shunt impedance $R_{sh}/m$  and quality factor Q parameters as function of the of $\frac{a}{\lambda}$ ratio is a polynomial of second degree. As a example, in case of $r_1/r_2$ = 5/7 elliptical shape, the Rsh/m and Q fit are given

\begin{equation}
R_{sh}/m = 293.3 - 856.37\  \frac{a}{\lambda}-170.05\  (\frac{a}{\lambda})^2 
\end{equation}

and 

\begin{equation}
Q = 5525.8-774.86\  \frac{a}{\lambda}+ 13548\ (\frac{a}{\lambda})^2.
\end{equation}

From the beam dynamics considerations and with satisfactory RF parameters, the analysis is restricted in the (0.09-0.2) $\frac{a}{\lambda}$ ratio range. 
From Figs. 5, 6, 7 for a given $\frac{a}{\lambda}$ ratio, for the longitudinal shunt impedance, $R_{sh}/m$, the unloaded quality factor Q and the coupling coefficient K have comparable values.

\begin{figure}[H]
\begin{minipage}{15pc}
\includegraphics[width=17pc]{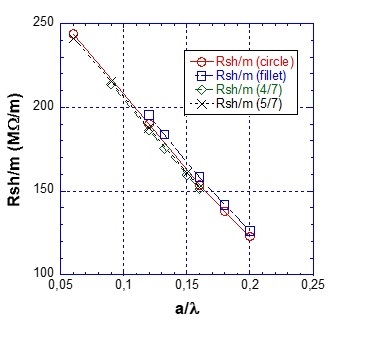}
\caption{\small{Shunt impedance as function iris aperture a/$\lambda$ ratio.}}
\end{minipage}\hspace{2pc}%
\begin{minipage}{15pc}
\includegraphics[width=17pc]{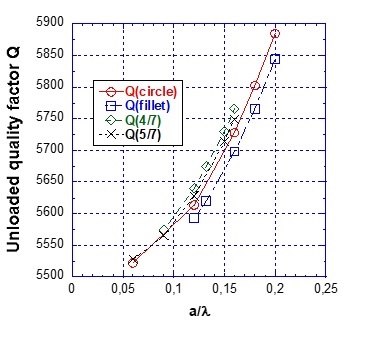}
\caption{\small{Shunt impedance as function iris aperture a/$\lambda$ ratio.}}
\end{minipage}\\
\begin{center}
\begin{minipage}{15pc}
\includegraphics[width=17pc]{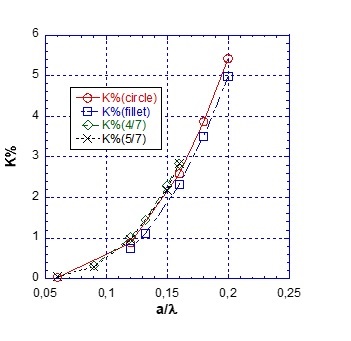}
\caption{\small{Coupling coefficient K ($\%$) as function of the a/$\lambda$ ratio.}}
\end{minipage}
\end{center}
\end{figure}

In the (0.1-0.15) $\frac{a}{\lambda}$ ratio range, these plots indicate that, the $TM_{010}$ mode $R_{sh}/m$, Q and K values vary respectively within the 23 $\%$, 1.9 $\%$ and 90 $\%$. In all cases, these values are reasonably feasible. We also observe that the parameter K is very sensitive by increasing the $a/\lambda$ ratio since the cavity bandwidth becomes larger as it is expected to be.  

From Figure 8, the $E_p/E_a$ ratio of the peak surface electric field to effective accelerating field for the elliptical iris shape has best results than fillet ones in the (0.06 - 0.12) a/$\lambda$ range. Then, in the (0.12-0.2) a/$\lambda$ range, with the $r_1/r_2$ = 4/7 ellipse shape, $E_p/E_a$  has worse result. In case of a $r_1/r_2$ = 5/7 ellipse semi-axes iris shape, at a/$\lambda$ = 0.12  (i.e. 1 mm aperture iris radius), $E_p/E_a$  is about 25 $\%$ less than the fillet one. From Figure 9, the $H_p/E_a$ ratio of magnetic field to effective accelerating field, differs at most by 1$\%$. By comparing the results of the elliptical semi-axes iris at 1 mm iris radius, the $H_p/E_a$ worse case is about 0.5 $\%$ larger than the fillet one. The variation of the cavity radius as function of the iris radius for different iris shape at the operating frequency is shown in Figure 10 at the operating frequency. For a given iris radius, the cavity radii have values comparable to each other as they are expected to be.

\begin{figure}[H]
\begin{minipage}{15pc}
\includegraphics[width=17pc]{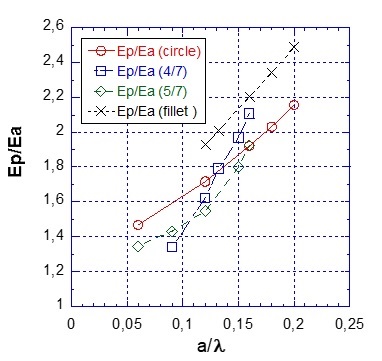}
\caption{\small{$E_p/E_a$ Cavity radius as function of the $a/\lambda$ ratio.}}
\end{minipage}\hspace{2pc}%
\begin{minipage}{15pc}
\includegraphics[width=17pc]{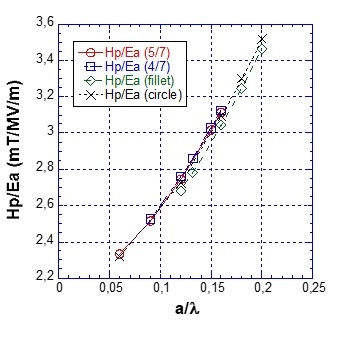}
\caption{\small{$H_p/E_a$ Cavity radius as function of the $a/\lambda$ ratio.}}
\end{minipage}\hspace{2pc}%
\end{figure}

As a compromise between the beam dimensions and the RF parameters, we choose a geometry with $\frac{a}{\lambda}$= 0.12 (or a=1mm radius), b = 3.628 mm, h = 0.667 mm and $r_1/r_2$ = 5/7. This choice allows to get a minimum  $E_p/E_a$ ratio and it does affect much the $H_p/E_a$ ratio. The longitudinal shunt impedance $R_{sh}/m$ = 188 M$\Omega$/m, the unloaded quality factor is Q = 5640 and the coupling coefficient K=1 $\%$.  

For sake of completeness, Figure 11 reports the dispersion curve of the proposed cavity estimated with the Superfish software by confirming the standard behaviour of the frequency $TM_{010}$ mode as function of the phase advance per cell.

\begin{figure}[H]
\begin{minipage}{15pc}
\includegraphics[width=17pc]{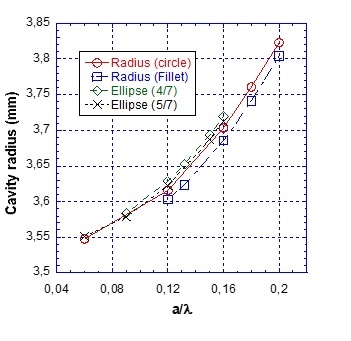}
\caption{\small{Cavity radius as function iris radius by keeping the operating frequency at F = 35.982 GHz.}}
\end{minipage}\hspace{1pc}%
\begin{minipage}{15pc}
\includegraphics[width=18pc]{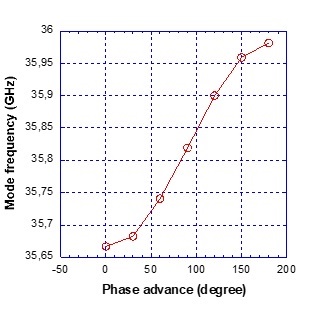}
\caption{\small{Dispersion relation of the SW structure.}}
\end{minipage}
\end{figure}

\section{Breakdown rate limit }

Intensive high power experimental activities on  single-cell SW structures were done at Stanford Linear Accelerator Centre (SLAC) in order to study the mechanism of breakdown phenomenon. The aim of the study was to obtain the maximum possible gradient for normal conducting accelerators and explore the limitation for higher gradient operation \cite{ref22,ref23,ref24}. As a result, the fundamental mechanism of the  vacuum RF breakdown is a complicated phenomenon which strongly limits the higher gradient accelerator performances. 
Usually, the cavity geometry has to be optimized  in order to minimize the peak surface electric field and to get a satisfactory surface magnetic field value. But a practical tolerable limit on the higher gradient operation came out recently from the experimental activity \cite{ref9,ref10,ref11} by estimating the Modified Poynting Vector and pulse heating (PH) effects. 
The Breakdown Rate (BDR) is a measure of the RF sparks per unit time and length inside an accelerating structure. A new quantity has been introduced in \cite{ref9,ref10}, which depends on the RF pulse length, the Modified Poynting Vector (MPV) defined as $S_c$=Re(S)+$\frac{1}{6}$Im(S) where S is the Poynting vector, in order to have a parameter to refer to during the linac design.

The values of the MPV and pulse heating (PH) as function of both circular and with a 5/7 elliptical shapes for different accelerating gradients by assuming a/$\lambda$ = 0.12 (or a = 1mm),  is reported in Table 2. The results have been obtained with the code HFSS.

\begin{table}[H]

\caption{Ellipse shape has $r_1/r_2$ = 5/7 semi-axes ratio}

\begin{center}

\small{ \begin{tabular}{|| c| c| c|c||}
\hline
Iris Geometry& $E_0$ [MV/m] &MPV \tiny{[MW/mm$^2$]} &PH [Celsius $^\circ$C] \\ 
 \hline\hline
Ellipse& 100 & 1.63& 7.79\\ 
\hline 
Ellipse&132 & 2.83 & 13.58\\ 
\hline
Ellipse & 150& 3.23& 17.67\\ 
\hline
Circle&100& 2.97& 6.97 \\ 
\hline
Circle& 132& 5.18 & 12.14 \\ 
\hline
 Circle & 150& 6.70& 15.69  \\ 
\hline
\end{tabular}}
\end{center}
\end{table}

The safety threshold of the MPV is estimated to be 5 $MW/mm^2$ at 100 ns RF pulse length \cite{ref9,ref10} and pulse heating should be below 50 $^\circ C$ \cite{ref11}. From the Table 2, we observe that  in all cases the safety thresholds of the Modified Poynting Vector and the pulse heating are satisfied. For sake of completeness, it should be noted that for the circular iris at 150 MV/m accelerating gradient, the Modified Poynting Vector could be at the limit of the allow safety threshold. 

For the SW Ka-Band structure, in case of rounded circle iris and by assuming a a/$\lambda$ = 0.12, we estimated a Modified Poynting Vector of $S_c$ $\approx$ 2.97 MW/$mm^2$ (well below the safety threshold of  6.3 MW/$mm^2$) for an accelerating gradient of $E_{acc}$=100 MV/m, and RF pulse length (flat top) of 50ns  \cite{ref9,ref10}. The RF pulsed heating  is estimated to be about $\Delta$T $\approx$ 6.97 $^\circ$C also well below the safety threshold \cite{ref11}. In case of elliptical iris with $r_1/r_2$ = 5/7, the Modified Poynting Vector is estimated to be of 1.63  MW/mm$^2$ and pulse heating is of about $\Delta$T  $\approx$ 7.79 $^\circ$C. With respect to the former rounded shape iris, the Poyinting vector value is improved by about 45 $\%$ while the pulse heating effect is worsened by about 10.5 $\%$. Both are still well below the safety thresholds and they are in agreement with the behaviour reported in Figs. (8) and (9)  and  at a/$\lambda$=0.12 obtained with SuperFish computing software.

As a result, we consider as best design the case of elliptical iris with $r_1/r_2$= 5/7. This case allows also to increase the accelerating gradient up to 130 MV/m since the Modified Poynting Vector is 2.83 MW/$mm^2$ and the pulse heating is $\Delta$T = 14 $^\circ$C. Both values give a good safety margin during the operation. Figures (12 a) and (12 b) illustrate the magnetic field magnitude and the Modified Poynting Vector calculated with the HFSS computing software with the parameters reported in Table 1.

\begin{figure}[H]
 \begin{center}
\includegraphics[width=0.45\linewidth]{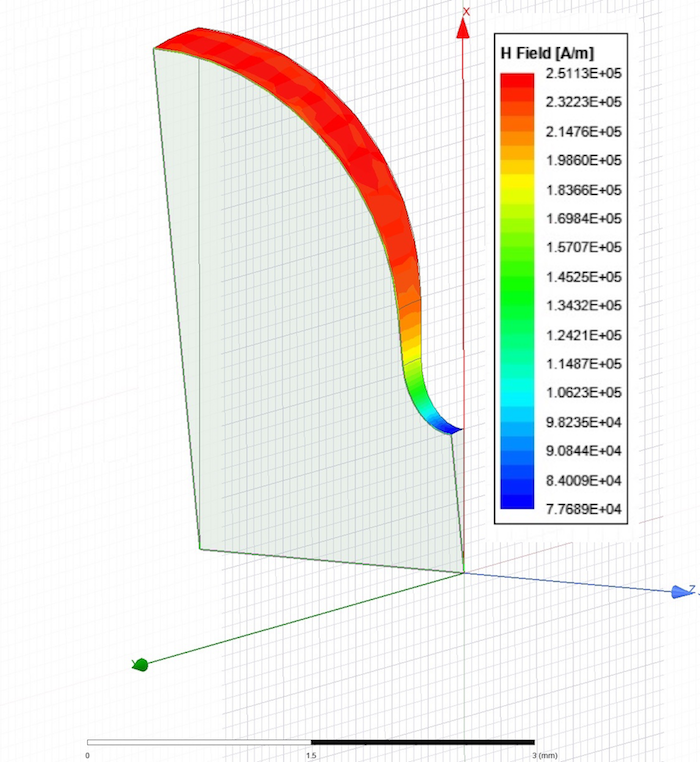}(a)
\includegraphics[width=0.45\linewidth]{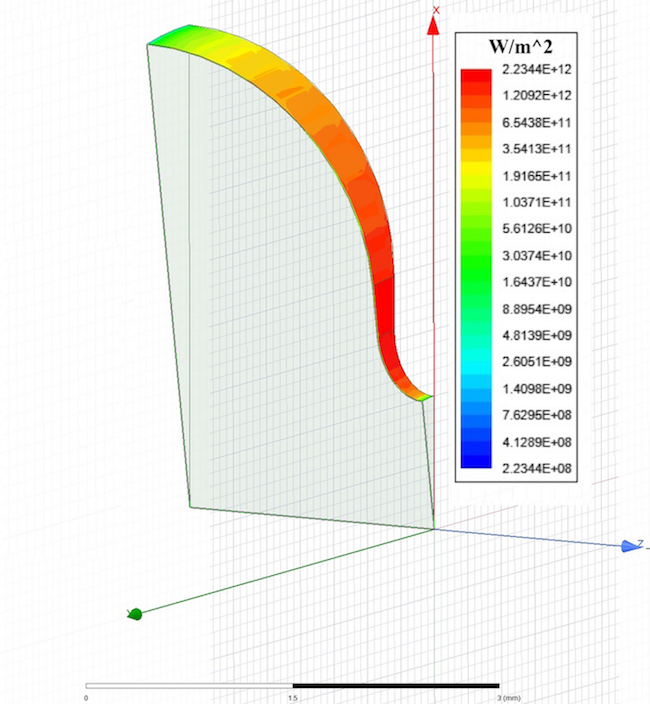}(b)

\caption { a) Magnetic field magnitude b) Modified Poynting Vector for the $TM_{01}$ mode of the high accelerating periodic structure.  }
\end{center}

      \end{figure}

Assuming a matched input power of 5 MW, 8 cm structure length, and cell length of 4.166 mm, we are able to obtain an effective accelerating field of 108 MV/m. Of course, a 36 GHz circulator,  for providing at least  2 MW peak power at F = 36 GHz, and a SLED system in order to get a 8 MW peak power for being able to feed the structure, have to be foreseen.
The change of resonance frequency as function of cavity, iris radius and cavity length, has been estimated to be around $\Delta$f = 11.1 MHz/$\mu$m, $\Delta$f = 2.4 MHz/$\mu$m, and $\Delta$f = 0.46 MHz/$\mu$m, respectively. By adjusting the cavity radius and the iris radius in opposite directions, the corresponding frequency shift is estimated to be of 13.5 MHz/$\mu$m. To summarize, the cavity frequency shift per unit radius can be expressed as  $\Sigma_{i=1}^3\ (\frac{\Delta f}{\delta x})_i$ (where i=1 refers to the cavity radius, i=2 to the iris radius adjustment and i=3 the cavity length ). As a result, the temperature tuning approach has to be foreseen because it is not possible to get a $\pm$ 1 $\mu m$ machining tolerance  as on the other hand it has been done for the 100 GHz structure \cite{ref25}.
The performance of the accelerating structure could also be limited by the resonant electron discharges or "multipactoring". According to our experience, it is well known that for reducing or eliminating this phenomenon it is recommended to have a curved profile of the cavity surfaces for "one point multipactoring" \cite{ref12}. Due to the big aperture of the structure, we believe that the "two points multipactoring" in the gap region of the structure is unlikely to occur due to presence of the greater magnetic force, thereby no resonant discharge can be determined \cite{PV-B.Spataro}. As a conclusion, we expect to have no particular problem on the multipactoring discharge for the chosen rounded cavity shape and large iris radius.

\section{Thermal and Stress Analysis}

A rise in temperature will vary the accelerator dimensions and the characteristic frequency  will change accordingly. The temperature rise can be reduced by means of a cooling system. For getting the frequency shift behaviour as a function of the temperature change, the thermal study is also required. We want to estimate the frequency shift caused by a change in temperature over the accelerating structure operating on $\pi$ mode. We will assume that a closed cooling water system is used in order to keep the operating temperature at  40 $^\circ$C.

The preliminary thermal and stress analysis was also carried out in CST. In Fig. 13, we show the result of the single cell where a cooling system with longitudinal pipes is assumed. The simulation is performed assuming a gradient of $E_{acc}$=125 MV/m with a corresponding average power per unit length of about 2 kW/m, with a water flux of 3 l/min. The hot spot is about 40 $^\circ$C (standard operation) and it can be lowered by adjusting water flux and water temperature. The consequent stress analysis shows a yield strength (Von Mises) $<$ 20 MPa which is below the safety threshold for copper ($\sim$ 70 MPa). The corresponding maximum displacement is about 1 $\mu$m (i.e. frequency shift is negligible or tunable). The cooling system will be optimized during final engineering (water jacket or brazed channels) in order to avoid water-to-vacuum leaks. 

\begin{figure}[H]
 \begin{center}
\includegraphics[width=0.4\linewidth]{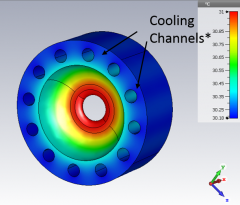}(a)
\includegraphics[width=0.4\linewidth]{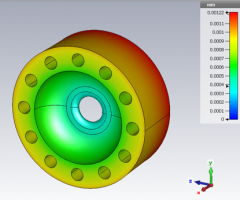}(b)

\caption { a) Thermal Simulation, single cell (35.982 GHz)  b) Stress Analysis (Von Mises), single cell (35.982 GHz) }
\end{center}

      \end{figure}

\section{Machining}

It has been experimentally demonstrated that hard copper is able to stand ultra-high gradients unlike the high-temperature treated one \cite{ref26,ref27} . As a result, we plan to machine the Ka-Band SW structure for high gradient applications in two halves with TIG welding of the outer surfaces \cite{ref1} and \cite{ref28}. 

\section{Conclusions}

In order to linearize the longitudinal space phase of the Compact Light XLS project, we have chosen a SW Ka-Band ($\sim$35 GHz) accelerating structure with elliptical iris with $r_1/r_2$=5/7 semi-axes ratio, by operating on $\pi$ mode at the third RF harmonic with respect to the main linac RF frequency ($\sim$11.99 GHz) at a 100 MV/m accelerating gradient. This structure can also work with a higher accelerating gradients up to 150 MV/m by keeping the Modified Poynting Vector and the pulse heating effects well below the safety tresholds. We are planning to finalize the structure design as well as engineering of the RF power source that will be able to produce up to a 5 MW input power by using a SLED system  \cite{ref3,ref4,ref5,ref6,ref7}. The dimensions of the cavity are perfectly consistent with the 100 GHz structures by scaling law with the frequency already tested at SLAC \cite{ref25,ref31}.  	
 In case of the single bunch operation, also a numerical and analytical study of the longitudinal and transverse wake-fields on the beam dynamic effects have  been carried out and discussed at first XLS Compact  Annual meeting in Barcelona Spain \cite{ref16}, and as a result, their estimation on the beam dynamics gave no specific trouble.

\section*{Acknowledgment}
This work was partially supported by INFN National Committee V through the ARYA Project and the Compact Light XLS Project, grant agreement N. 777431.





\end{document}